\documentclass{pasj02}
\Received{$\langle$reception date$\rangle$}
\Accepted{$\langle$acception date$\rangle$}
\Published{$\langle$publication date$\rangle$}

\newcommand{\sgr}{Sgr~A~East}
\newcommand{\axj}{AX~J$1745.6-2901$}
\newcommand{\hea}{He$\alpha$}
\newcommand{\lya}{Ly$\alpha$}

\usepackage{amsmath}
\usepackage[switch,mathlines]{lineno}
\usepackage{url}

\begin{document}

\title{Overionized plasma in the supernova remnant Sagittarius A East anchored by XRISM observations}

\author{XRISM Collaboration\altaffilmark{1}}
\author{Marc Audard\altaffilmark{2}}
\author{Hisamitsu Awaki\altaffilmark{3}}
\author{Ralf Ballhausen\altaffilmark{4,5,6}}
\author{Aya Bamba\altaffilmark{7}}
\author{Ehud Behar\altaffilmark{8}}
\author{Rozenn Boissay-Malaquin\altaffilmark{9,5,6}}
\author{Laura Brenneman\altaffilmark{10}}
\author{Gregory V.\ Brown\altaffilmark{11}}
\author{Lia Corrales\altaffilmark{12}}
% ...

\altaffiltext{1}{\textit{Corresponding Authors: Hiromasa Suzuki, Hideki Uchiyama, Masayoshi Nobukawa, and Yuki Amano}}
\altaffiltext{2}{Department of Astronomy, University of Geneva, Versoix CH-1290, Switzerland} 
\altaffiltext{3}{Department of Physics, Ehime University, Ehime 790-8577, Japan} 
\altaffiltext{4}{Department of Astronomy, University of Maryland, College Park, MD 20742, USA} 
\altaffiltext{5}{NASA / Goddard Space Flight Center, Greenbelt, MD 20771, USA} 
\altaffiltext{6}{Center for Research and Exploration in Space Science and Technology, NASA / GSFC (CRESST II), Greenbelt, MD 20771, USA} 
\altaffiltext{7}{Department of Physics, University of Tokyo, Tokyo 113-0033, Japan} 
\altaffiltext{8}{Department of Physics, Technion, Technion City, Haifa 3200003, Israel}
\altaffiltext{9}{Center for Space Science and Technology, University of Maryland, Baltimore County (UMBC), Baltimore, MD 21250, USA}
\altaffiltext{10}{Center for Astrophysics | Harvard-Smithsonian, MA 02138, USA} 
\altaffiltext{11}{Lawrence Livermore National Laboratory, CA 94550, USA} 
\altaffiltext{12}{Department of Astronomy, University of Michigan, MI 48109, USA} 
% ...

\email{hiromasa050701@gmail.com}


\maketitle

\begin{abstract}
Sagittarius A East is a supernova remnant with a unique surrounding environment, as it is located in the immediate vicinity of the supermassive black hole at the Galactic center, Sagittarius A$^*$.
The X-ray emission of the remnant is suspected to show features of overionized plasma, which would require peculiar evolutionary paths.
We report on the first observation of Sagittarius A East with X-Ray Imaging and Spectroscopy Mission (XRISM). {Equipped with a combination of} high-resolution microcalorimeter spectrometer and large field-of-view CCD imager, we for the first time resolved the Fe \emissiontype{XXV} K-shell lines into fine structure lines and measured the forbidden-to-resonance intensity ratio to be $1.39 \pm 0.12$, which strongly suggests the presence of overionized plasma.
We obtained a reliable constraint on the ionization temperature
just before the transition into the overionization state, to be $> 4$~keV. The recombination timescale was constrained to be $< 8\times10^{11}$~cm$^{-3}$~s.
The small velocity dispersion of $109 \pm 6$~km~s$^{-1}$ indicates a low Fe ion temperature $< 8$~keV {and a small expansion velocity $< 200$~km~s$^{-1}$.}
%
The high initial ionization temperature and small recombination timescale suggest that {either rapid cooling of the plasma via adiabatic expansion from dense circumstellar material or intense photoionization by Sagittarius A$^*$ in the past may have triggered the overionization.
}
\end{abstract}



\section{Introduction}
Supernova remnants usually begin their lives with very low ionization states and evolve toward the collisional ionization equilibrium (CIE).
{Over the last} decade, however, X-ray observations have disclosed that certain fraction of the remnants contain ``overionized'' plasma, in which ionization degrees are significantly higher than those expected if the plasma is ionizing or in CIE (see \cite{yamaguchi20} for review).
When and how they transitioned to overionzation states still {remain under debate}.
{Diagnosing overionized plasma in supernova remnants is important because it would hold the key to understanding the possible variety in the evolutionary paths.}

Sagittarius A East ({\sgr}) is a supernova remnant of great interest because of its location of the immediate vicinity of the supermassive black hole Sagittarius A$^*$ (Sgr A$^*$) and suspected presence of overionized plasma.
The angular size of the radio shell $\approx 3.5' \times 2.5'$ \citep{ekers75} corresponds to the physical size of $\approx 8~{\rm pc} \times 6~{\rm pc}$ assuming a distance of $8$~kpc \citep{reid93, gillessen09, frail11, ranasinghe22}.
The age of the remnant is poorly constrained from the dynamics and non-equilibrium plasma state to be $10^{3\text{--}4}$~yr \citep{maeda02, fryer06, koyama07}.
%

%
Its center-filled X-ray structure, {with the shell-like radio emission, indicates that it is classified into the ``mixed-morphology'' class \citep{rho98}.
As the process to realize this morphology is still unclear and the remnants in this class are often associated with overionized plasma, X-ray spectroscopy is of great importance to diagnose the plasma state and its history.}
The X-ray spectrum shows intense Fe K-shell emission lines, indicating the presence of a high-temperature plasma \citep{maeda02, sakano04, park05, koyama07, ono19, zhou21}. The plasma state is still under debate. \citet{ono19} claims the detection of overionized plasma based on the Fe-K radiative recombination continuum feature, whereas \citet{zhou21} found no strong evidence for overionized plasma.
Considering its surrounding environment, it would be reasonable to suspect that the plasma state could have been affected by hypothetical past activities of Sgr A$^*$ \citep{ryu09, ryu13, nakashima13, ono19}.
High-resolution X-ray spectroscopy is thus desired to resolve the plasma state and its origin.

We report on the first high-resolution spectroscopy of {\sgr} with XRISM (X-Ray Imaging and Spectroscopy Mission; \cite{tashiro18, tashiro20}).
In this work, we focus on narrow-band spectroscopy of the prominent Fe \emissiontype{XXV} (He-like) and Fe \emissiontype{XXVI} (H-like) K-shell emission lines {(hearafter {\hea} and {\lya}, respectively)}, to provide a robust constraint on the plasma state as well as to demonstrate the diagnostic power of XRISM.

\section{Observations and data reduction}\label{sec-obs}
XRISM is equipped with the microcalorimeter Resolve \citep{ishisaki18} and wide field-of-view (FoV) CCD camera Xtend \citep{mori22}.
Resolve consists of 35 active pixels, covering a sky region $\approx 0\farcm5 \times 0\farcm5$ each and $\approx 3\farcm1 \times 3\farcm1$ combined, with an energy resolution of $< 5$~eV (FWHM: full width half maximum; \cite{xrism24_n132d}).
Xtend is composed of four CCDs, giving a $\approx 38'\times38'$ FoV with an energy resolution of $< 200$~eV in FWHM. The angular resolution is $< 1\farcm7$ (HPD: half power diameter).

The {\sgr} and a nearby-sky regions were observed with XRISM from Feb. 26 to 29, 2024 UT (ObsID: 300044010) and Feb. 29 to Mar. 3 UT (ObsID: 300045010), respectively.
We reduce the data with the pre-release Build 7 XRISM software with HEAsoft ver. 6.32 \citep{heasarc14} and calibration database (CALDB) ver. 8 {(v20240815) \citep{terada21, loewen20}.}
We exclude periods of the Earth eclipse and sunlit Earth's limb, and South Atlantic Anomaly passages.
The effective exposures of Resolve (Xtend) left after the standard data reduction are 126~ks (108~ks) and 74~ks (63~ks) for the {\sgr} and nearby-sky observations, respectively. 
The systematic uncertainty in the energy scale of Resolve is evaluated using the onboard $^{55}$Fe source and found to be very small, $\approx 0.1$~eV at 5.9~keV {\citep{eckart24, porter24}.}

In our spectral analysis, we use XSPEC ver. 12.13.1 \citep{arnaud96} with the $C$-statistic \citep{cash79}, and AtomDB ver. 3.0.9 \citep{foster12} with the PyAtomDB package \citep{foster20}.
In calculation of charge exchange X-ray emission, we partly use the CX model \citep{gu16} in SPEX ver. 3.08.00 \citep{kaastra96}.
Redistribution matrix files (RMFs) are generated with the {\tt rslmkrmf} task using the cleaned event file and CALDB based on ground measurements. Line-spread function components include the Gaussian core, exponential tail to low energy, escape peaks, and Si fluorescence.
Auxiliary response files (ARFs) are generated with the {\tt xaarfgen} task. The assumed emission profiles for individual spectral components are described in Section~\ref{sec-spec}. 
Errors quoted in the text, figures, and tables indicate $1\sigma$ confidence intervals.

\begin{figure*}[htb!]
    \centering
    \includegraphics[width=16cm]{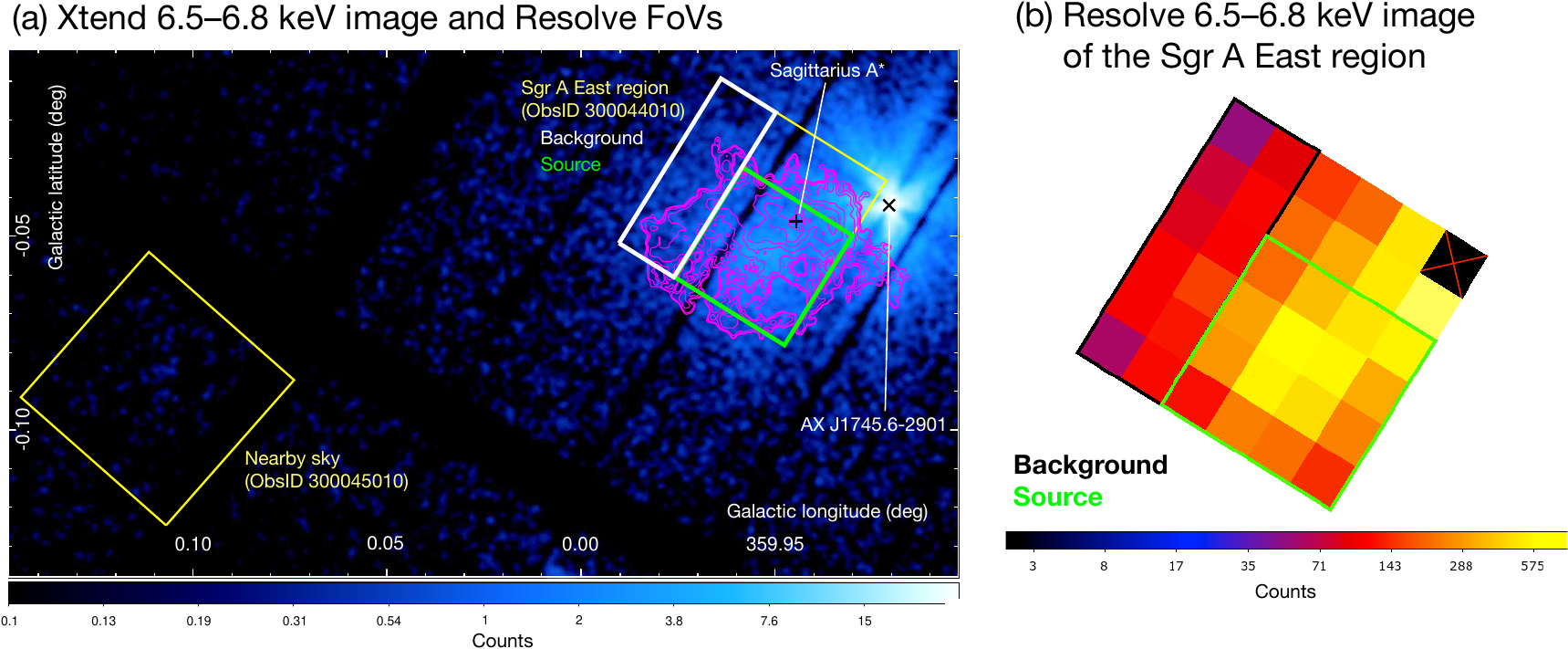}
    \flushleft
    \caption{(a) 6.5--6.8~keV Xtend image of the Galactic center region with the Resolve field of views (FoVs; yellow boxes), source (green) and background (white) regions. {Background subtraction and vignetting (off-axis exposure) correction is not applied.} The VLA 4.8~GHz contours {obtained from the NRAO archive (\url{https://www.vla.nrao.edu/astro/archive/pipeline/position/J174535.6-285839/})} are overlaid in magenta. (b) 6.5--6.8~keV Resolve image of the {\sgr} region with the source (green) and background (black) regions.
    \label{fig-image}}
\end{figure*}

\section{Analysis and results}\label{sec-ana}

\begin{figure*}[htb!]
    \centering
    \includegraphics[width=16cm]{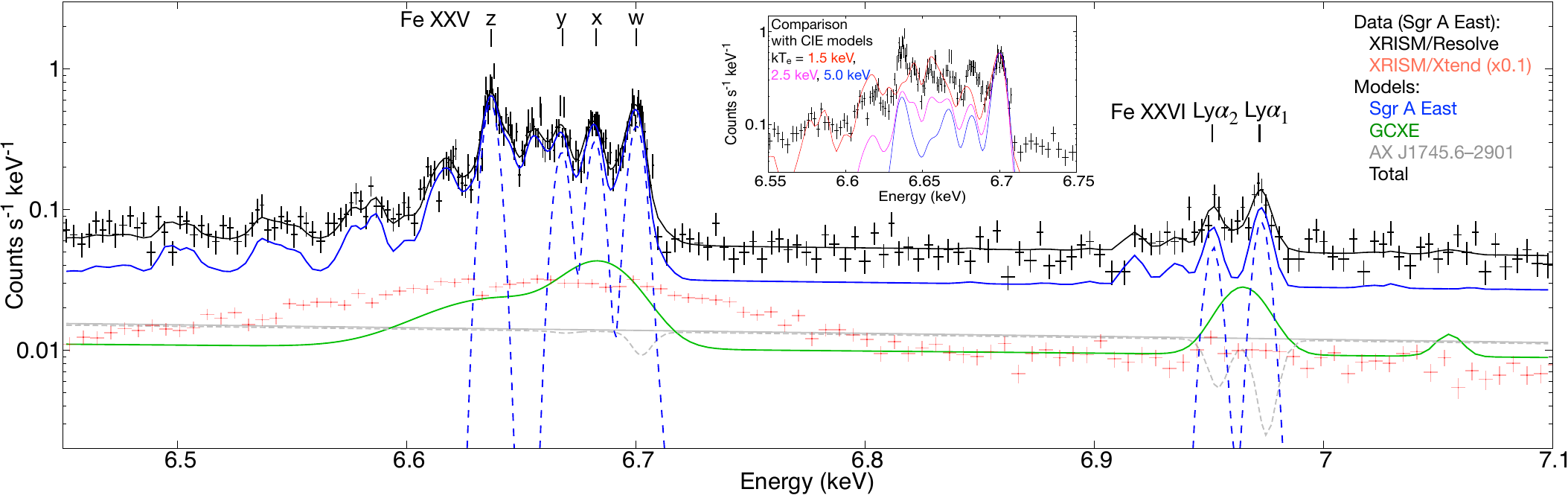}
    \flushleft
    \caption{X-ray spectrum of the {\sgr} region with XRISM Resolve with the best-fit spectral models. The blue solid line represents the {\sgr} model. The blue dashed lines indicate the Lorentzian models which replace the Fe-{\hea} and {\lya} lines in the \texttt{bvvrnei} model (see text for methodology). The solid and dashed gray lines represent the spectral models of {\axj} without and with absorption by ionized Fe atoms, respectively. A reference spectrum with {XRISM Xtend} is overplotted in red. The inset shows the collisional ionization equilibrium (CIE) plasma model with different electron temperatures ($kT_{\rm e}$) and a comparison with the data. 
    \label{fig-spec}}
\end{figure*}

\subsection{Spectral modeling of {\sgr}}\label{sec-spec}
Figure \ref{fig-image} shows 6.45--6.8~keV (Fe {\hea}) Xtend and Resolve images extracted from the observation of {\sgr}.
One can see that the Resolve observation of {\sgr} is highly affected by the bright transient source ({low-mass X-ray binary}), {\axj} \citep{hyodo09}, {whose X-ray emission has a black-body continuum with absorption lines at Fe {\hea} and {\lya} and thus requires a careful treatment. Further assessment can be found later in this section.}
The Resolve spectrum extracted from the source region is shown in Figure~\ref{fig-spec}, demonstrating that the Fe {\hea} triplet is successfully separated for the first time.

Our primary interest is the Fe {\hea} and {\lya} line intensities.
{We perform plasma diagnostics based on these intensities measured with a semi-phenomenological model rather than simply relying on plasma models.
This treatment is to minimize the potential bias due to the other spectral features including the high and uncertain background.
}
%
The Resolve spectrum clearly exhibits evidence for an overionized plasma, i.e., enhanced Fe {\hea} z/w and x/y ratios and coexistence of Fe ions with a broad range of charge states (H-like, He-like, and lower-ionized ones like Li-like and Be-like).
{We apply Lorentzian models with a Gaussian-like broadening to obtain the individual fluxes of the Fe {\hea}-z, y, x, w, {\lya}$_1$, and {\lya}$_2$ (six in total).}
{To explain the other spectral features including weak lines and continua, we apply a two-temperature overionized plasma model (\texttt{bvvrnei + bvvrnei} in XSPEC) plus a power-law function \citep{ono19} without the Fe {\hea} and {\lya} lines (see \cite{suzuki20d} for the methodology).}\footnote{{The origin of the power-law component is thought to be either an ensemble of point sources or non-thermal emission associated with filaments in the remnant \citep{koyama07}, or many faint X-ray reflection nebulae \citep{ono19}.}}
%
The Lorentzian widths (FWHM) are fixed to the individual natural widths. The Gaussian-like broadening (same for all the six lines in velocity dispersion) is tied to that of the \texttt{bvvrnei} model.
All the parameters of the low-temperature component are fixed to the values in \citet{ono19} except for the emission measure, the ratio of which relative to the high-temperature component is fixed to the value in \citet{ono19}. {The contribution of the low-temperature component at Fe {\hea} and {\lya} is below 5\% \citep{ono19} and thus its uncertainty does not affect the discussion below.}
The power-law index is fixed to 1.0 \citep{ono19}.
The electron temperature $kT_{\rm e}$, recombination timescale $\tau$, emission measure, velocity dispersion, and redshift of the high-temperature component, and power-law flux are treated as free parameters. The initial ionization temperature $kT_{\rm init}$ is fixed to 10~keV \citep{ono19} because the model itself is insensitive to it without the six major lines.
The interstellar absorption column, which is unimportant in the energy range we use, is fixed to $1.5\times10^{23}$~cm$^{-2}$ (based on \cite{ono19}, roughly consistent with \cite{zhou21}).
As the input emission profile to generate ARFs, we use an Fe {\hea} image obtained with Chandra.

The background of our observation is dominated by the Galactic center X-ray emission (GCXE; see \cite{koyama18} for review) and {\axj}.
%
We determine the spectral model for {\axj} using an Xtend spectrum (ObsID: 300044010) extracted from a circle with a $1'$ radius centered at (R.A., Dec.) = (266\fdg3985, $-29\fdg0261$). The emission is approximated with a simple blackbody spectrum. We first ignore the absorption features at Fe {\hea} and {\lya} \citep{trueba22}, which are time-variable. The uncertainty due to this treatment is evaluated later in this section.
When we apply this model to the Resolve spectra, we only allow the normalization to vary within $\pm\, 50\%$ to account for uncertainties in the {ray-tracing software when generating ARF}, with the other parameters fixed.
Meanwhile, we empirically model the spectrum extracted from the FoV of the nearby-sky observation (ObsID: 300045010) to obtain the GCXE spectral shape.\footnote{Details of the GCXE spectrum itself will be reported in a separate paper.}
Only the surface brightness is left as a free parameter, which is assumed to be the same for the source and background regions. The effect of the possible spatial dependence is evaluated later in this section \citep{muno04, chat15}.
As the emission profiles to be inputted to ARFs, we assume a point-like and flat sources for the {\axj} and GCXE, respectively.

We simultaneously fit the source and background spectra with the combined model, {\sgr} + GCXE + {\axj}, {taking into account the mutual contamination of each component between the two regions}.
We allow the normalization of the {\sgr} model for the background spectrum to vary within $\pm\, 50\%$ relative to that for the source region to account for uncertainties in the {ray-tracing software and thus in the ARF}.
As a result, we obtain an acceptable fit with a $C$-stat/d.o.f. $= 2975.1/2589$ (Figure~\ref{fig-spec}). The obtained parameters are summarized in Table~\ref{tab-spec}.
We successfully constrain the intensity ratios as Fe {\hea}-z/{\hea}-w $= 1.39 \pm 0.12$,  {\lya}/{\hea}-w $= 0.36 \pm 0.07$.
The best-fit {\lya}$_1$/{\lya}$_2$ intensity ratio is consistent with $\approx 2$, which is generally expected in thermal plasma.
The constrained velocity dispersion, $\approx 109$~km~s$^{-1}$, corresponds to an Fe ion temperature of $\approx 8$~keV if the broadening is purely due to thermal motion.
%
One may expect to detect a strong radiative recombination edge of He-like Fe ions at $\approx 8.83$~keV as was the case for W49B \citep{ozawa09}. Although our model indeed reproduces a similar feature because of the spectral parameters similar to W49B (e.g., \cite{yamaguchi18}), the feature is much less noticeable because of its {much lower flux than the contamination from the power-law component of {\sgr}, {\axj}, and GCXE.}


\begin{table}[htb!]
    \caption{Results of the spectral modeling for {\sgr}}
    \centering
    \begin{tabular}{ll}
        \hline\hline
        Parameter & Value \\
        \hline
        $N_{\rm H}$ ($10^{22}$~cm$^{-2}$) & 15 (fixed) \\
        Broadening (km~s$^{-1}$) & $109 \pm 6$ \\
        Redshift & $(2 \pm 5)\times10^{-5}$ \\
        \multicolumn{2}{l}{Lorentzian flux ($10^{-5}$ ph s$^{-1}$ cm$^{-2}$)}  \\
        ~~~{\hea} z (6636.58~eV) & $3.19 \pm 0.17$  \\
        ~~~{\hea} y (6667.55~eV) & $1.43 \pm 0.14$  \\        
        ~~~{\hea} x (6682.30~eV) & $1.71 \pm 0.14$  \\        
        ~~~{\hea} w (6700.40~eV) & $2.30 \pm 0.16$  \\
        ~~~Ly$\alpha_1$ (6973.07~eV)  & $0.49 \pm 0.08$  \\
        ~~~Ly$\alpha_2$ (6951.86~eV) & $0.33 \pm 0.07$  \\
        \multicolumn{2}{l}{High-$kT_{\rm e}$ {\tt bvvrnei} model$^{*}$}\\
        ~~~$kT_{\rm e}$ (keV) & $1.69 \pm 0.09$  \\
        ~~~$kT_{\rm init}$ (keV) & 10 (fixed) \\
        ~~~$\tau$ ($10^{11}$~s~cm$^{-3}$) & $(7.7 \pm 1.1) \times 10^{11}$ \\
        ~~~Fe (solar) & 1.5 (fixed) \\
        ~~~${\it n}_\mathrm{e} {\it n}_\mathrm{H} V$ (cm$^{-5})^\dag$ & $(5.0 \pm 0.5)\times10^{58}$ \\
        Power-law flux$^{\ddag}$ & $(1.8 \pm 0.7)\times 10^{-13}$ \\
        GCXE flux$^{\S}$ & $(3.0 \pm 0.1) \times 10^{-14}$ \\
         \hline
    \end{tabular}
    \label{tab-spec}
    \flushleft
    {*} {Note that the {\tt bvvrnei} model is introduced to explain the lines except for the prominent Fe {\hea} and {\lya}, and continua.}\\
    {\dag} Emission measure, where {\it n}$_\mathrm{e}$ and {\it n}$_\mathrm{H}$ stand for electron and hydrogen number densities (cm$^{-3}$). A distance of 8~kpc is assumed. \\
    {\ddag} {Absorbed} photon flux in 6.5--7.1~keV (erg~s$^{-1}$~cm$^{-2}$).\\
    {\S} {Absorbed} photon flux in 6.5--7.1~keV (erg~s$^{-1}$~cm$^{-2} \, {\rm arcmin}^{-2}$).
\end{table}

We evaluate the contribution of two processes which possibly alter the Fe {\hea} forbidden-to-resonance ratio, i.e., the Fe {\hea} resonance scattering and charge exchange (CX) X-ray emission. Details are described in Appendix~\ref{sec-rscx}.
The possible flux decrease of Fe {\hea}-w due to scattering is estimated to be $\approx 5\%$ at most. The possible contribution of the CX emission is also evaluated to be $< 10\%$ to the Fe {\hea}-z/{\hea}-w and {\lya}/{\hea}-w ratios. Thus, both will not affect the plasma parameters significantly.

To account for systematic uncertainties in the background estimation, we consider uncertainties associated with the GCXE flux and Fe-K absorption features of {\axj}. 
%
Based on the Fe {\hea} flux distribution over a $17' \times 17'$ region around Sgr A$^{*}$ \citep{muno04} and the projected stellar mass ratio of the source and background regions \citep{chat15}, 
we take into account the possible spatial dependence of GCXE by scaling the best-fit GCXE flux of the source region by $+40\%$ and search for the best-fit parameters again.
The Fe {\hea} and {\lya} fluxes are slightly changed but within $\approx 10\%$, both smaller than the statistical errors and are insignificant.
We evaluate the effect of the absorption features of {\axj} by replacing the model with a one with the K-shell absorption lines of ionized Fe determined with the Xtend spectrum and confirmed with the Resolve spectrum extracted from westmost pixels (dashed gray line in Figure~\ref{fig-spec})\footnote{Details of the spectral features of {\axj} will be reported in a separate paper.}.
The resultant line fluxes of {\sgr} are slightly increased, by $\approx 18\%$ at Fe {\lya}, yielding insignificant changes in the plasma parameters.

\subsection{Plasma diagnostics based on Fe-{\hea} and {\lya}}\label{sec-diag}

\begin{figure*}[htb!]
    \centering
    \includegraphics[width=14cm]{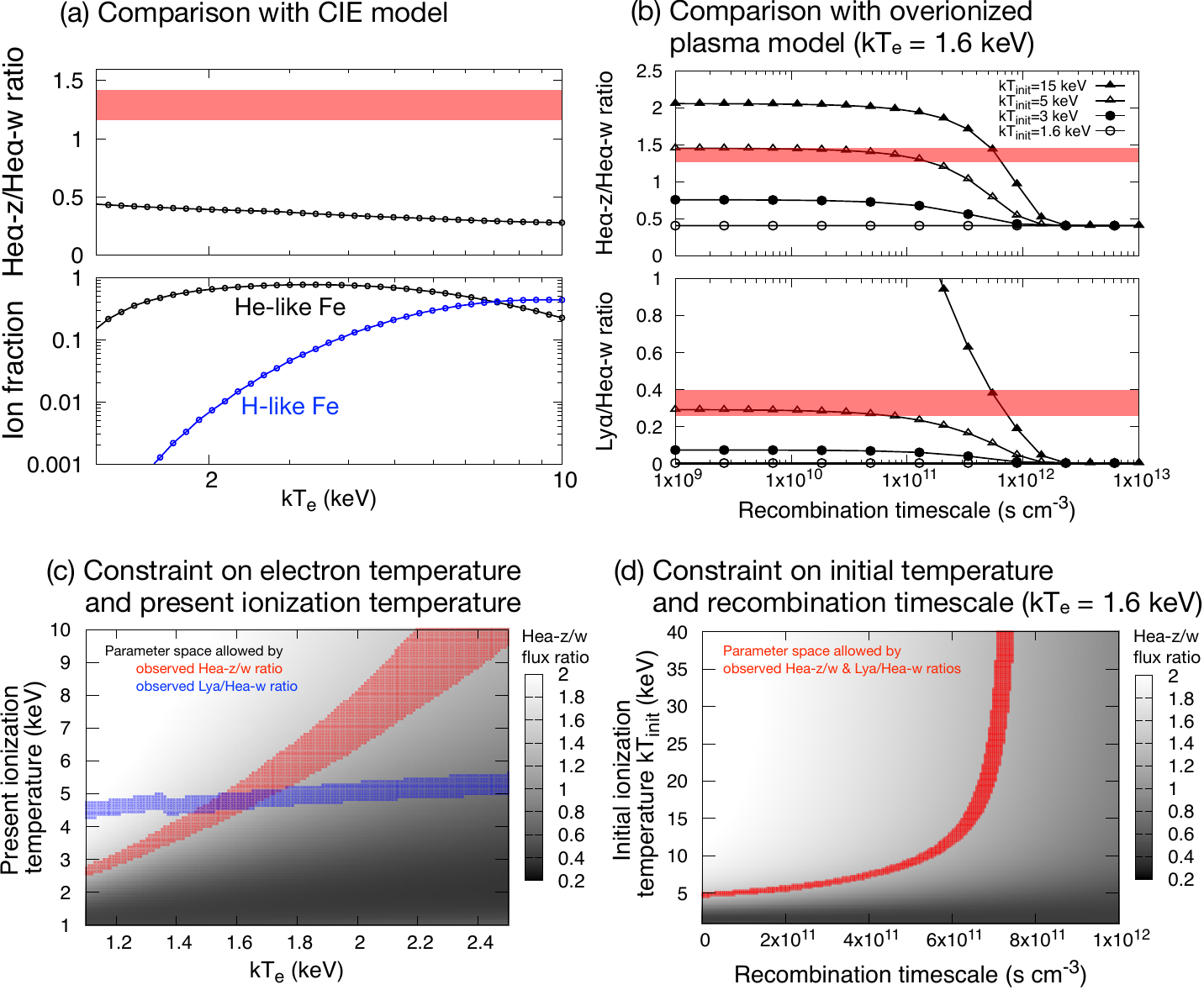}
    \flushleft
    \caption{Comparison of the observed Fe {\hea}-z/{\hea}-w and {\lya}/{\hea}-w intensity ratios and AtomDB plasma models. The observed intensity ratios and their $1\sigma$ uncertainty ranges are indicated with red transparent regions in the panels (a) and (b). (a) Comparison with the CIE model with various electron temperatures ($kT_{\rm e}$). The upper and lower panels show the {\hea}-z/{\hea}-w intensity ratio and ion fractions, respectively. (b) Comparison with the overionized plasma model (\texttt{rnei} with $kT_{\rm e} = 1.6$~keV) with various initial temperatures ($kT_{\rm init}$). The upper and lower panels show the {\hea}-z/{\hea}-w and {\lya}/{\hea}-w ratios as a function of recombination timescale, respectively. (c) $1\sigma$ uncertainty ranges of $kT_{\rm e}$ and present ionization temperature which satisfy the observed {\hea}-z/{\hea}-w (red) and {\lya}/{\hea}-w (blue) ratios. (d) $1\sigma$ confidence region of the recombination timescale and initial ionization temperature ($kT_{\rm init}$) (red) which satisfy both the observed {\hea}-z/{\hea}-w and {\lya}/{\hea}-w ratios. The background color scale in (c) and (d) shows the {\hea}-z/{\hea}-w intensity ratio.
    \label{fig-diag}}
\end{figure*}

We compare the determined line intensities with the plasma models in AtomDB to constrain the plasma state.
Figure~\ref{fig-diag} (a) indicates that the CIE plasma model can not explain the observed high forbidden-to-resonance intensity ratio with a reasonable range of electron temperature, where a sufficient amount of both He- and H-like Fe atoms are present.
Figure~\ref{fig-diag} (b) shows that the overionized plasma model can explain the observed line intensity ratios with sufficiently high initial ionization temperatures $kT_{\rm init}$.
%
Figure~\ref{fig-diag} (c) and (d) quantify the constrained ranges of the electron temperature, present and initial ionization temperatures, and recombination timescale based on the Fe {\hea}-z/{\hea}-w and {\lya}/{\hea}-w intensity ratios.
The electron temperature $kT_{\rm e}$ and present ionization temperature $kT_{\rm z}$\footnote{CIE temperature equivalent to the ionization state which explains the observed line ratios {(e.g., \cite{masai94})}.}  are estimated to be $kT_{\rm e} = 1.6 \pm 0.2$~keV and $kT_{\rm z} = 4.7 \pm 0.4$~keV, respectively. The initial ionization temperature is thus constrained to be $kT_{\rm init} > 4$~keV.
Note that no $kT_{\rm init}$ values can explain the observations when $\tau > 8\times10^{11}$~s~cm$^{-3}$, as such timescales lead to much lower ionization states than those observed.

{The electron temperature and recombination timescale directly constrained from the Fe {\hea} and {\lya} line ratios are consistent with those shown in Table~\ref{tab-spec}. This fact means that the plasma state required from the Fe {\hea} and {\lya} lines can explain other spectral features including the satellite lines as well.
Thus, we apply a model without replacing the Fe {\hea} and {\lya} lines with Lorentzian functions
for {\sgr} and indeed obtain a very similar fit, with $C$-stat/d.o.f. = 2976.1/2595.}

\section{Discussion}\label{sec-dis}
\subsection{Basic properties of {\sgr}}\label{sec-basic}
{The derived recombination timescale, $\tau <8\times10^{11}$~s~cm$^{-3}$, can be used to evaluate the age of {\sgr}. If we assume that the Fe-K emitting plasma is dominated by ejecta, which is reasonable given its center-filled distribution and over abundance \citep{koyama07, ono19}, we can derive the plasma density: typical CSM + ejecta masses of $\gtrsim 1$~solar with the radius {of the Fe {\hea} emission traced with Chandra}, 1.6~pc yield a plasma density $n_{\rm e} \gtrsim {1}$~cm$^{-3}$ {for both solar-abundance or pure-Fe plasma}, which are similar to the estimates from the plasma emission measures in previous studies \citep{maeda02, sakano04, koyama07, ono19} and this work. The age of the remnant should be larger than the elapsed time after the overionization occurred, $\tau /n_{\rm e} \approx 2500~{\rm yr}\, (\tau/8\times10^{11}~{\rm s}~{\rm cm}^{-3})(n_{\rm e}/10~{\rm cm}^{-3})^{-1}$. This will be more meaningful once some additional constraints are obtained.}

%
{The velocity dispersion of $\approx 109$~km~s$^{-1}$ is converted to an upper-limit expansion velocity of $\approx 200$~km~s$^{-1}$ \citep{xrism24_n132d}.
If we assume that the Fe-K emitting plasma is dominated by ejecta, expansion velocities will be kept high, such as $\gtrsim 1000$~km~s$^{-1}$, even after shock-heating (e.g., \cite{sato17, xrism24_n132d}). Thus, the small expansion velocities inferred in {\sgr} will require certain processes to suppress the expansion. The inferred high ambient density is qualitatively consistent in this sense. Alternatively, if the progenitor wind produced dense circumstellar material (CSM), it would naturally cause a suppression of the expansion.
}
The low Fe ion temperature $< 8$~keV, {along with the high electron temperature $\approx 1.6$~keV, generally requires a long thermal relaxation timescale $\gtrsim 10^{11}$~s~cm$^{-3}$ and a moderate shock velocity $\lesssim 2000$~km~s$^{-1}$ in the unshocked ejecta frame, given non-equilibrium of the temperatures of different elements immediately behind shocks \citep{xrism24_n132d, ohshiro24}.
We note that this may not hold for {\sgr} because the evolutionary path may have been unusual, e.g., the remnant may have experienced a quick equilibration behind the shock due to dense CSM as we discuss in Section~\ref{sec-over}.
}
%

The upper limit of the observed redshift of Fe {\hea} is converted to a line-of-sight velocity of $\approx 20$~km~s$^{-1}$. This supports the idea that {\sgr} is very close to Sgr A$^*$: line-of-sight component of the Galactic rotation velocity, $< 50$--150~km~s$^{-1}$ away from the solar system at a distance of 1--10~pc from Sgr A$^*$ \citep{lugten86}, may be observed as $< 20$~km~s$^{-1}$ {after applying the standard corrections for the line-of-sight velocities of the solar system ($\approx 10$~km~s$^{-1}$) and Earth ($\approx 28$~km~s$^{-1}$).
This estimate is also similar to the velocities of molecular clouds in the vicinity ($\sim 50$~km~s$^{-1}$: e.g., \cite{tsuboi99, tsuboi11}).}

\subsection{Origin of the overionization}\label{sec-over}
{Here we discuss how the overionized plasma in {\sgr} was produced.}
We briefly examine three candidate cases {which are widely considered} as the origin of the overionization in supernova remnants, rapid cooling of the plasma via (a) thermal conduction with cold material {\citep{kawasaki02}} or (b) adiabatic expansion from dense CSM {\citep{itoh89}}, and (c) enhanced ionization due to irradiation of charged particles or photons {(e.g., \cite{ono19})}.
In the cases (a) and (b), the ionization temperature $kT_{\rm z}$ before a transition to overionization states is similar to or lower than the electron temperature $kT_{\rm e}$ assuming an ordinary evolution. A rapid cooling of electrons causes a transition and realizes $kT_{\rm z} > kT_{\rm e}$.
In the case (c), the electron temperature is kept the same before and after the transition, and only the ionization state is quickly enhanced and realizes $kT_{\rm z} > kT_{\rm e}$.

If we assume that the thermal conduction (a) was the origin of the overionization in {\sgr}, the electron temperature before the transition should have been $> 4$~keV.
It is hard to realize such high electron temperatures in supernova remnants in general. In fact, electron temperatures this high have never been observed in remnants, although may be possible theoretically in such a dense environment, where the plasma would quickly reach CIE.
%
The thermal conduction scenario would not naturally explain such high initial temperatures due to the expected ordinary evolution before the transition, and thus is less favorable.

The adiabatic expansion scenario (b) may satisfy the high initial ionization temperature because it assumes the existence of dense CSM, developing a strong reverse shock at an early stage ($\sim 100$~yr; \cite{itoh89}). The resultant higher shock velocity and higher density than without dense CSM in combination quickly achieve a high ionization state.
Indeed, X-ray observations performed within $\sim 100$~days of several supernovae exploding in dense CSM suggested the presence of plasma with high electron temperatures $> 10$~keV (e.g., \cite{chandra24}).
%
Since the CSM breakout should occur within a few 100~yr, the recombination timescale should be similar to or larger than the ``age times present plasma density''.
{Thus, if the age of the remnant is smaller than $\approx 25000~{\rm yr}\, (n_{\rm e}/1~{\rm cm}^{-3})^{-1}$, this scenario can explain the recombination timescale measured in our spectroscopy. Better constraints on the age is desired to test this scenario more precisely.}


Realizing sufficiently high ionization states to explain the observed overionized plasma by enhanced ionization (c) will be hard in general in the case of collisional ionization with charged particles \citep{sawada24}. Another possibility is photoionization by a luminous X-ray source. Having such a high luminosity source in the vicinity of a remnant is rare and the photoionization scenario has never been considered as a plausible origin for overionized remnants (e.g., \cite{kawasaki02}).
In the case of {\sgr}, however, intense photoionization may be possible because of the presence of Sgr A$^*$ in the immediate vicinity.
As already suggested by \citet{ono19}, a luminosity $L \gtrsim {10^{40}}$~erg~s$^{-1} \,(n_{\rm e}/1~\text{cm}^{-3})(R/1~\text{pc})^2$, with the $R$ being the distance from Sgr A*, is required to realize an ionization state equivalent to the CIE with 5~keV ($\log\xi \approx 3$).
This lies in between those of the hypothetical outbursts of Sgr A$^*$, $L \sim 10^{39}$~erg~s$^{-1}$ a few 100 yr ago \citep{ryu09, ryu13} and $L \sim 10^{44}$~erg~s$^{-1}$ $\sim 10^{5\text{--}6}$~yr ago \citep{su10, nakashima13}.
Thus, if a past outburst of Sgr A$^*$ with $L \gtrsim {10^{40}}$~erg~s$^{-1}$ occurred $\lesssim 2500~{\rm yr}\, (n_{\rm e}/10~{\rm cm}^{-3})^{-1}$ ago, it would have initiated the overionization of {\sgr}.
We note that, we would expect overionization of other elements in the swept-up interstellar medium as well in this scenario. A work on the broad-band spectrum, which is important in this sense, will be reported separately.

\section*{Acknowledgement}
Part of this work was support by JSPS KAKENHI grant numbers 24K17093, 22H00158, 22H01268, 22K03624, 23H04899, 23K03454, 19K21884, 20H01947, 20KK0071, 23K20239, 24K00672, 21K13963, 24K00638, 21K13958, 24K17104, 20K04009, 21K03615, 24K00677, 20K14491, 23H00151, 19K14762, 23K03459, 24K17105, 24H00253, 23K13154, 21H01095, 23K20850, 21H04493, 20H01946, 20H05857, 22KJ1047,
%
NASA under award Nos. 80GSFC21M0002, 80NNSC22K1922, 80NSSC20K0733, 80NSSC20K0737, 80NSSC24K0678, 80NSSC23K0738, 80NSSC18K0978, 80NSSC20K0883, 
JSPS Core-to-Core Program (grant number:JPJSCCA20220002),
STFC through grant ST/T000244/1,
NASA contract NAS8-0360,
%
%
the U.S. Department of Energy by Lawrence Livermore National Laboratory under Contract DE-AC52-07NA27344,
%
%
the Kagoshima University postdoctoral research program (KU-DREAM),
the Alfred P. Sloan Foundation through the Sloan Research Fellowship,
%
%
%
the Strategic Research Center of Saitama University,
%
%
the Canadian Space Agency (grant 18XARMSTMA),
%
%
%
%
%
the RIKEN Pioneering Project Evolution of Matter in the Universe (r-EMU), Rikkyo University Special Fund for Research (Rikkyo SFR),
the Waterloo Centre for Astrophysics and generous funding to B.R.M. from the Canadian Space Agency and the National Science and Engineering Research Council of Canada,
and NSF award 2205918.

\begin{figure*}[htb!]
    \centering
    \includegraphics[width=16cm]{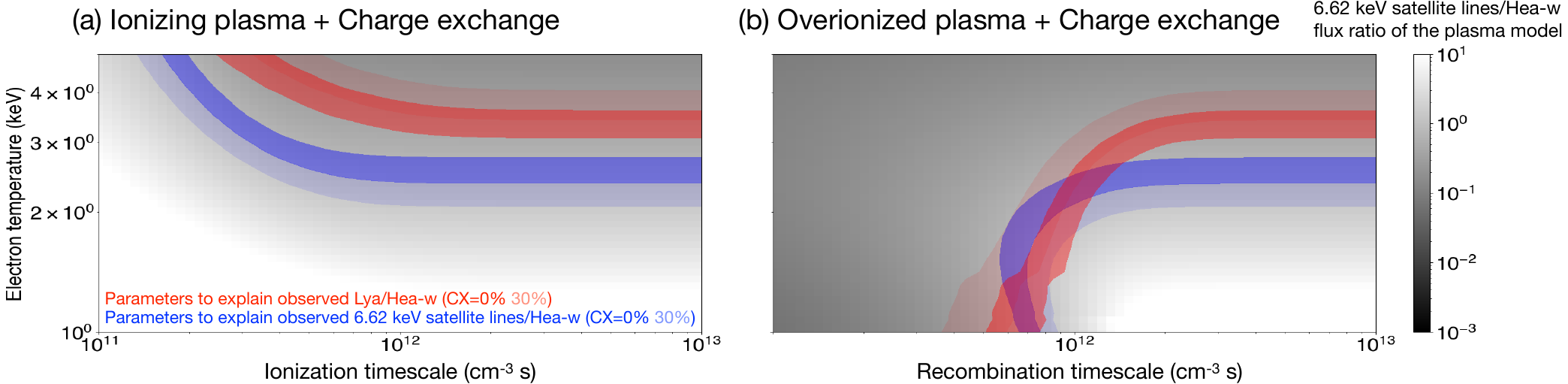}
    \flushleft
    \caption{Constraints on the parameters of the (a) ionizing and (b) overionized plasma models with the contribution of charge exchange (CX) emission (ACX2 v2.0 in XSPEC) considered. The initial ionization temperature of 10~keV is assumed for the overionized plasma model. The red and blue transparent regions show the observed intensity ratios of $\approx 6.62$~keV satellite lines/Fe {\hea}-w and Fe {\lya}/{\hea}-w, respectively. The regions with lower and higher transparencies indicate the CX flux levels of 0\% and 30\% at {\hea}-w with respect to the plasma models, respectively. 
    \label{fig-cx}}
\end{figure*}

\bibliography{references.bib}
\bibliographystyle{test_pasj.bst}

\appendix
\section{Contribution of resonance scattering and charge exchange emission}\label{sec-rscx}

We here evaluate the contribution of two processes which possibly alter the Fe Hea forbidden-to-resonance ratio, i.e., the resonance scattering and charge exchange (CX) X-ray emission.
Due to the high oscillator strength, Fe {\hea}-w (resonance) line may be scattered in the plasma itself depending on the optical depth and geometry, thereby reducing the line-of-sight intensity. {We use the formulation by \citet{kaastra95} to estimate the transmission probability $p = 1/(1 + 0.43\tau)$, with the $\tau$ being the optical depth of the plasma \citep{kastner90}}.\footnote{We assume reasonable or conservative parameters for {\sgr}, Fe abundance $= 1.5$ solar, electron temperature $= 1.6$~keV, recombination timescale $= 8\times 10^{11}$~s~cm$^{-3}$, initial temperature $= 10$~keV, without a velocity dispersion, electron density $= 20$~cm$^{-3}$, and line-of-sight length of 3.2~pc. We also assume that a photon completely escapes from the line of sight when scattered. The oscillator strength and ion fractions are taken from SPEX.}
The resultant probability of the scattering is $< 5\%$, indicating that the resonance scattering is negligible.

As for the CX emission due to an interaction with neutral material, we first examine if the ionizing plasma model with the contribution of CX can explain the observations based on ACX2 v2.0 in AtomDB\footnote{We assume the same Fe abundance, temperature, and ionization state for the CX component as those of the thermal plasma model. A collision velocity of 1000~km~s$^{-1}$ is assumed. Ions are assumed to repeatedly capture electrons until they become neutral.}. As a conservative approach, only the spectral shape of the CX model is considered and the absolute emission measure is not used in our evaluation. Figure~\ref{fig-cx} shows a data-to-model comparison of the $\approx 6.62$~keV satellite lines/{\hea}-w and {\lya}/{\hea}-w intensity ratios. One can see that the ``ionizing plasma + CX'' model cannot explain both the two line ratios at the same time, with an even increased discrepancy when the contribution of CX is higher. In the case of the ``overionized plasma + CX'', one can find a certain upper limit of the allowed contribution of CX.
We then evaluate the upper limit of its contribution to the {\hea}-z/{\hea}-w and {\lya}/{\hea}-w line ratios precisely using ACX2 (AtomDB) and CX (SPEX), obtaining $<10\%$ (95\% upper limit), which do not alter the derived plasma parameters significantly.

\end{document}